\documentclass[
aps,
prd,
floats,
floatfix,
twocolumn,
superscriptaddress,
nofootinbib,
noshowpacs]{revtex4-2}

\usepackage{hyperref}
\usepackage{amssymb}
\usepackage{amsmath}
\usepackage{physics}
\usepackage{latexsym}

\usepackage{booktabs}
\usepackage{multirow}

\usepackage{verbatim}
\usepackage{mathrsfs}
\usepackage{amsfonts}

\usepackage{graphicx,subfigure}
\usepackage{epsfig}

\usepackage{units}
\usepackage{overpic}
\usepackage[utf8]{inputenc}
\usepackage{rotating}
\usepackage{enumerate}

\usepackage{soul}

\usepackage{xfrac}
\usepackage{xcolor}
\usepackage{multirow}

\begin{document}


\title{The continuum spectrum of nonrelativistic multi-frequency Proca stars}

\author{Galo Diaz-Andrade}
\affiliation{Departamento de Física, División de Ciencias e Ingenierías, Campus León, Universidad de Guanajuato, 37150, León, México}
\author{Alberto Diez-Tejedor}
\affiliation{Departamento de Física, División de Ciencias e Ingenierías, Campus León, Universidad de Guanajuato, 37150, León, México}
\author{Jose Luis Medina-Garcia}
\affiliation{Departamento de Física, División de Ciencias e Ingenierías, Campus León, Universidad de Guanajuato, 37150, León, México}
\author{Armando A. Roque}
\affiliation{Unidad Acad\'emica de F\'isica, Universidad Aut\'onoma de Zacatecas, 98060 Zacatecas, M\'exico}


\date{\today}

\begin{abstract}

Multi-frequency Proca stars are excited selfgravitating solutions of the $s=1$ Schr\"odinger-Poisson system that generalize the conventional stationary states of a massive vector field.
 Unlike stationary states, which are characterized by a single oscillation frequency, multi-frequency configurations exhibit a quasi-periodic dynamics involving two or three distinct frequencies. In this paper, we present a systematic study of the spectrum of spherical multi-frequency Proca stars and show that, at fixed particle number, they form continuous families interpolating between discrete stationary states of constant linear polarization. Furthermore, we analyze their stability and demonstrate that a subset of these multi-frequency configurations are linearly stable against general perturbations. In particular, we show that a necessary, although not sufficient, condition for stability is the presence of a non-negligible nodeless component, and that radial stability alone is not sufficient to guarantee full linear stability. 
 Finally, we briefly discuss the potential implications of multi-frequency states for proving the particle spin in ultralight dark matter models.
\end{abstract}

\maketitle

\section{Introduction}

Proca stars are selfgravitating equilibrium configurations of spin-1 particles that occupy the same quantum state and can be described by a classical massive vector field.
They were first introduced by Brito, Cardoso, Herdeiro, and Radu in Ref.~\cite{Brito:2015pxa} as regular and asymptotically flat solutions of the classical Einstein-Proca equations, and have since been exhaustively studied in the literature~\cite{SalazarLandea:2016bys, Brihaye:2017inn, Sanchis-Gual:2017bhw, Minamitsuji:2018kof, Sanchis-Gual:2018oui, Herdeiro:2020kba, Herdeiro:2020jzx, CalderonBustillo:2020fyi, Herdeiro:2021lwl, Sanchis-Gual:2022mkk, Rosa:2022tfv, Herdeiro:2023wqf, Wang:2023tly,  Sengo:2024pwk}. In this work, we focus on the nonrelativistic limit of the theory, which is described by the classical $s=1$ Schr\"odinger-Poisson system~\cite{Jain:2021pnk, Zhang:2021xxa, Adshead:2021kvl,  Gorghetto:2022sue, Zhang:2023ktk, Chen:2024vgh, Zhang:2024bjo,Nambo:2024hao,Nambo:2025lnu}, and further restrict ourselves to spherically symmetric configurations.

The spectrum of nonrelativistic spherical Proca stars consists of {\it stationary} and {\it multi-frequency} states~\cite{Nambo:2024hao}. Stationary states are characterized by wave functions that evolve in time with one frequency, whereas in a multi-frequency state two or three distinct frequencies can be present simultaneously. Stationary states, including constant and radially polarized configurations, have been studied in~\cite{Jain:2021pnk, Zhang:2021xxa, Adshead:2021kvl}, and we concentrate here on multi-frequency states, which were first introduced in Ref.~\cite{Nambo:2024hao} and remain largely unexplored.

As shown in Ref.~\cite{Nambo:2024hao} through a simple example (see also~\cite{Nambo:2025lnu}), spherical 2-component multi-frequency states present a continuum spectrum that interpolates between stationary states of constant linear polarization. In this work, we extend this study by systematically exploring the solution space of spherical 2- and 3-component multi-frequency states, including an analysis of their linear stability. Interestingly, we show that a subset of these configurations, despite corresponding to excited states of the $s=1$ Schr\"odinger-Poisson system at fixed particle number, are nonetheless stable against small perturbations. This opens the possibility that multi-frequency states could form dynamically and coexist with the ground state, which sources the same gravitational field as standard boson stars and is therefore gravitationally indistinguishable from a configuration of spin-$0$ particles. In this case, the presence of additional frequencies would then provide a distinctive signature of particle spin in ultralight dark matter models~\cite{Marsh:2015xka,Hui:2021tkt,Ferreira:2020fam}. We discuss this possibility further at the end of the paper. The main results of this work are summarized in Table~\ref{table}.

\begin{table*}
\caption{{\bf Spherical equilibrium configurations.} Classification of the equilibrium solutions of the spherical $s=1$  Schr\"odinger-Poisson system at fixed particle number $N$ into {\it states}, {\it classes}, and {\it families}. Stationary states were previously studied in~\cite{Nambo:2024hao,Nambo:2025lnu}, while the present work focuses on multi-frequency states. 1-component multi-frequency states are equivalent to stationary states with constant linear polarization. More generally, 1- and 2-component multi-frequency states can be regarded as particular cases of 3-component multi-frequency states in which one or more components vanish. The definitions of the quantities listed in the table are provided in the main text.}
\begin{center}
\begin{tabular}{l l l l c l}
\toprule
State & Class & Family & Solution multiplicity & Frequencies & Stability \\
\midrule
\multirow{2}{*}{Stationary}
& Linear polarization 
& $(n)$ 
& One per family 
& 1
& Ground state $(0)$ \\

& Radial polarization 
& $(n)$ 
& One per family 
& 1
& Nodeless state $(0)$ \\
\midrule
\multirow{3}{*}{Multi-frequency}
& 1-component 
& $(n_x)$ 
& One per family 
& 1
& Ground state $(0)$ \\

& 2-component 
& $(n_x,n_y)$ 
& 1-parameter families 
& 2
& Stability bands in $(0,n_y)$ \\

& 3-component 
& $(n_x,n_y,n_z)$ 
& 2-parameter families 
& 3
& Stability bands in $(0,n_y,n_z)$ \\
\bottomrule
\end{tabular}
\end{center}
\label{table}
\end{table*}

\section{Multi-frequency Proca stars}

Our starting point is the $s=1$ Schr\"odinger-Poisson system,
\begin{subequations}\label{eqs.SP}
\begin{eqnarray}
i\frac{\partial\vec{\psi}}{\partial t} &=& -\frac{1}{2m_0}\Delta\vec{\psi}+m_{0}\mathcal{U}\vec{\psi}, \\
\Delta\mathcal{U} &=& 4\pi G m_0 n, 
\end{eqnarray}
\end{subequations}
where $\vec{\psi}(t,\vec{x})$ is a complex-valued vector wave function, $n(t,\vec{x}):=\vec{\psi}^*\cdot\vec{\psi}$ is the particle number density, and $\mathcal{U}(t,\vec{x})$ is the gravitational potential. The mass of the vector field is denoted by $m_0$, $G$ is Newton's constant, and we have set the reduced Planck constant to one, $\hbar=1$. Note that this system is invariant under the scaling transformation
\begin{equation}\label{eq.scaling.free}
t\mapsto\lambda_*^{-1}t, \quad 
\vec{x}\mapsto\lambda_*^{-1/2}\vec{x}, \quad 
\mathcal{U}\mapsto\lambda_* \mathcal{U},\quad \vec{\psi} \mapsto \lambda_*\vec{\psi},
\end{equation}
where $\lambda_*$ is an arbitrary positive and nonvanishing constant. 

\subsection{Spherical configurations}

In spherical symmetry, multi-frequency Proca stars are described by the following  ansatz~\cite{Nambo:2024hao}:
\begin{equation}\label{eq.ansatz.multi.cartesian}
\vec{\psi}(t,\vec{x})= \sum_{i=1}^3 e^{-iE_i t}\sigma_i^{(0)}(r)\hat{e}_i,
\end{equation}
where $E_i$ are constant real frequencies associated with the wave function components, $\sigma_i^{(0)}$ are real-valued functions depending only on the radial coordinate $r$, and $\hat{e}_i$ are the Cartesian unit basis vectors, with $i=x,y,z$ (or equivalently $i=1,2,3$). 

If we introduce the ansatz~(\ref{eq.ansatz.multi.cartesian}) into the $s=1$ Schr\"odinger-Poisson system~(\ref{eqs.SP}), we obtain 
\begin{subequations}\label{eqs.multi.frequency}
\begin{eqnarray}
E_i\sigma_i^{(0)}&=&\left(-\frac{1}{2m_0}\Delta_s + m_0\mathcal{U}\right)\sigma^{(0)}_i,\label{SEc2.2.2.2}\\
\Delta_s\mathcal{U} &=& 4\pi Gm_0 n^{(0)}, \label{s=1GPP.stationary.3}
\end{eqnarray}
\end{subequations}
where $\Delta_s:=\frac{1}{r}\frac{d^2}{dr^2}r$ denotes the radial Laplace operator and $n^{(0)}=\sum_i|\sigma_i^{(0)}|^2$ is the particle number density. To fully specify the solutions of Eqs.~(\ref{eqs.multi.frequency}), this system must be complemented with  appropriate boundary conditions. Regularity at the origin, $r=0$, requires
\begin{subequations}\label{Eq.BounCond.multy.1}
\begin{eqnarray}
\sigma_i^{(0)}(r=0)&=\sigma_{i0},\quad \sigma_i^{(0)\prime}(r=0)=0,\\
\mathcal{U}(r=0)&=\mathcal{U}_{0},\hspace{0.75cm} \mathcal{U}'(r=0)=0,
\end{eqnarray}
where $\sigma_{i0}$ and $\mathcal{U}_0$ are constants. At spatial infinity, $r\to\infty$, we impose
\begin{equation}
\lim\limits_{r\to\infty}\sigma^{(0)}_i(r)=0,\quad \lim\limits_{r\to\infty}\mathcal{U}(r)= 0.
\end{equation}
\end{subequations}
The asymptotic condition on the components of the wave function guarantees finite energy solutions, while the condition on the gravitational potential  is imposed by convention.

Equivalently, Eqs.~(\ref{eqs.multi.frequency}) can be expressed as an integro-differential nonlinear equation
\begin{equation}\label{Eq:SPA_NLsystem}
	E_i \sigma_{i}^{(0)}= \hat{\mathcal{H}}[n^{(0)}]\sigma_{i}^{(0)},
\end{equation}
with $\hat{\mathcal{H}}[n^{(0)}]$ a $\vec{\psi}$-dependent Hamiltonian operator defined by 
\begin{align}\label{eq.def.H}
\hat{\mathcal{H}}[n^{(0)}] := -\frac{1}{2m_0}\Delta_s + 4\pi Gm_0^2\Delta_s^{-1}(n^{(0)}).
\end{align}
Here, $\Delta_s^{-1}$ denotes the inverse of $\Delta_s$, defined as
\begin{equation}
[\Delta^{-1}_s(f)] (r) :=-\int_0^\infty \frac{f(\tilde{r})}{r_>} \tilde{r}^2 d\tilde{r},\label{LaplInve}
\end{equation}
where $f$ is an arbitrary function that depends only on the radial coordinate  $r$, and where we have set $r_{>}:=\max \left\lbrace r, \tilde{r} \right\rbrace$.

Notice that the system~(\ref{eqs.multi.frequency}) [or equivalently, Eq.~(\ref{Eq:SPA_NLsystem})], together with the boundary conditions~(\ref{Eq.BounCond.multy.1}), define a nonlinear eigenvalue problem for the frequencies $E_i$. For each set of amplitudes $(\sigma_{x0}, \sigma_{y0}, \sigma_{z0})$, there exists an infinite, countable set of solutions 
\begin{equation}
(\hat{E},\vec{\sigma}^{(0)
})=(E_{x,n_x}, E_{y,n_y}, E_{z,n_z}, \sigma_{x,n_x}^{(0)}, \sigma_{y,n_y}^{(0)}, \sigma_{z,n_z}^{(0)}),
\end{equation}
one for each $(n_x,n_y,n_z)$, where $n_x, n_y$, and $n_z = 0, 1, 2, \ldots$ represent the number of nodes of the functions $\sigma_{x,n_x}^{(0)}(r), \sigma_{y,n_y}^{(0)}(r)$, and $\sigma_{z,n_z}^{(0)}(r)$ in the interval $0 < r < \infty$, respectively. Without loss of generality, we will restrict our analysis to  configurations satisfying $n_x < n_y < n_z$.\footnote{In principle, the values of $n_x$, $n_y$ and $n_z$ can be any non-negative integers. However, according to the nodal theorem, two components $\sigma_{i}^{(0)}(r)$ and $\sigma_{j}^{(0)}(r)$ with $i\neq j$ and the same node numbers are proportional to each other (see, e.g., Sec.~IV B in  Ref.~\cite{Nambo:2024hao} for details). Therefore, configurations with $n_i = n_j$ do not yield new solutions. Moreover, we can always order the integers such that $n_x<n_y<n_z$.}

The invariance of Eqs.~(\ref{eqs.SP}) under global internal $U(3)$ transformations acting on the field $\vec{\psi}$ implies the conservation of a Hermitian second-rank tensor $\hat{Q}$. For each solution $(\hat{E},\vec{\sigma}^{(0)})$, this tensor is diagonal, with components 
\begin{equation}
Q_{ij}=N_i\delta_{ij},
\end{equation}
where $N_i$ represents the particle number associated with each component $\sigma^{(0)}_i(r)$,
\begin{equation}
N_i = 4\pi  \int_0^\infty \sigma^{(0)2}_i r^2 dr.\end{equation}
These quantities provide a natural way to describe and compare multi-frequency states. The total number of particles in a configuration is then given by $N = \sum_i N_i$.
In addition, multi-frequency states have total energy
\begin{equation}
\mathcal{E} = -4\pi \sum_i \int_0^\infty \frac{1}{2m_0}\sigma_i^{(0)\prime2}r^2 dr,
\end{equation}
and possess vanishing spin, $\vec{S}=0$, and orbital, $\vec{L}=0$, angular momentum, so that their total angular momentum $\vec{J}=\vec{S}+\vec{L}=0$ also vanishes.

To provide a systematic exploration of the solution space of multi-frequency Proca stars, we organize them into classes and families. We first distinguish three different classes, according to the number of independent frequencies with which the vector field oscillates in internal space. This number can be chosen to coincide with the number of nonvanishing components of the wave function present in the configuration (see footnote~\ref{footnote} in Sec.~\ref{sec.1component} for an explanation). Accordingly, we identify 1-component, 2-component, and 3-component multi-frequency states.\footnote{1-component multi-frequency Proca stars oscillate with a single frequency and therefore reduce to stationary Proca stars. We nevertheless include them for completeness, so as to provide a unified framework for all configurations.}

Within each class, we further organize the solutions into families, labeled by the ordered set $(n_x,n_y,n_z)$, where $n_i$ denotes the node number of the corresponding component. The number of entries in this set equals the number of nonvanishing components: for 1-component states we write $(n_x)$, for 2-component states $(n_x,n_y)$, and for 3-component states $(n_x,n_y,n_z)$. For a given family $(n_x,n_y,n_z)$, individual configurations are specified by the central amplitudes $(\sigma_{x0}, \sigma_{y0}, \sigma_{z0})$, which can be varied continuously while keeping the total particle number 
$N$ fixed. Moreover, configurations with different values of $N$ are related through the scaling symmetry in Eq.~(\ref{eq.scaling.free}), which implies that a single choice of $N$ is sufficient to characterize the full solution space within each family.
A schematic summary of this classification into states, classes, and families is provided in Table~\ref{table}.

\subsection{Linear perturbations}

In practice, we are interested in solutions that are linearly stable. For this purpose, we analyze the evolution of linear perturbations, which we parametrize as:
\begin{equation}
\vec{\psi} (t,\vec{x})= e^{-i \hat{E} t}\left[\vec{\sigma}^{(0)}(\vec{x}) 
 + \epsilon\vec{\sigma}(t,\vec{x})+\mathcal{O}(\epsilon^2) \right],
\label{eq:ansatzPert}
\end{equation}
where $(\hat{E}, \vec{\sigma}^{(0)})$ is a multi-frequency solution  and $\vec{\sigma}(t, \vec{x})$ is a complex vector-valued function describing the perturbation to first order in the small parameter $\epsilon$.

The derivation of the linearized system was presented in detail in Ref.~\cite{Nambo:2025lnu}, where it was shown that, for multi-frequency states, the perturbations satisfy the equations
\begin{subequations}\label{Eq:Perturbation.multi}
\begin{align}
    i \lambda \vec{\mathcal{A}} &= \left[\hat{\mathcal{H}}^{(0)} - \hat{E}\right] \vec{\mathcal{B}} ,\\
    i \lambda \vec{\mathcal{B}} &= \left[\hat{\mathcal{H}}^{(0)} - \hat{E}\right] \vec{\mathcal{A}} + 16\pi G m_0^2\Delta_s^{-1}\left( \vec{\sigma}^{(0)} \cdot \vec{\mathcal{A}}\right) \vec{\sigma}^{(0)},
\end{align}
\end{subequations}
where the function $\vec{\sigma}(t,\vec{x})$ has been decomposed as
\begin{equation}
\label{Eq:PertAnsatz}
\vec{\sigma}(t,\vec{x}) = \left[ \vec{\mathcal{A}}(\vec{x})+\vec{\mathcal{B}}(\vec{x}) \right]e^{\lambda t} + \left[\vec{\mathcal{A}}(\vec{x})-\vec{\mathcal{B}}(\vec{x})\right]^{*}e^{\lambda^* t}.
\end{equation}
Here $\vec{\mathcal{A}}$ and $\vec{\mathcal{B}}$ are two complex vector-valued functions of $\vec{x}$, $\lambda$ is the associated complex eigenvalue, and $\hat{\mathcal{H}}^{(0)}:=\hat{\mathcal{H}}[n^{(0)}]$ was previously introduced in Eq.~(\ref{eq.def.H}).

Expanding the perturbations $\vec{\mathcal{A}}(\vec{x})$ and $\vec{\mathcal{B}}(\vec{x})$ in terms of the spherical harmonics as
\begin{subequations}\label{Eq:CartesianPert}
\begin{eqnarray}
\vec{\mathcal{A}}(\vec{x}) &=& \sum_i\sum_{JM} A_{JM}^i(r)Y^{JM}(\vartheta,\varphi)\hat{e}_i ,\\
\vec{\mathcal{B}}(\vec{x}) &=& \sum_i\sum_{JM} B_{JM}^i(r)Y^{JM}(\vartheta,\varphi)\hat{e}_i ,
\end{eqnarray}
\end{subequations}
where $A^i_{JM}$ and $B^i_{JM}$ are complex-valued radial functions, and after some algebra, one obtains a system of equations that can be written in the matrix form
\begin{equation}\label{eq.general system}
i\lambda \left( \begin{array}{c}
X_{JM} \\ Y_{JM} \\ Z_{JM} \end{array} \right) = \left( \begin{array}{ccc}
 M_{JM}^{11} & M_{JM}^{12} & M_{JM}^{13}  \\
 M_{JM}^{21} & M_{JM}^{22} & M_{JM}^{23}  \\ 
 M_{JM}^{31} & M_{JM}^{32} & M_{JM}^{33}  \\
\end{array} \right)\left( \begin{array}{c}
X_{JM} \\ Y_{JM} \\ Z_{JM} \\ \end{array} \right),
\end{equation}
where $J$ and $M$ denote the angular momentum and magnetic quantum numbers, respectively, with $J=0,1,2,\ldots$ and $M=-J,\ldots, J$. The variables $X_{JM}$, $Y_{JM}$, and $Z_{JM}$, as well as the matrix $M^{ij}_{JM}$ (which depends on the background equilibrium configuration), are defined as
\begin{equation}\label{eq.defXYZ1}
X_{JM} = \left( \begin{array}{c}
A_{JM}^x \\ B_{JM}^x \end{array} \right), \,
Y_{JM} = \left( \begin{array}{c}
A_{JM}^y \\ B_{JM}^y \end{array} \right),\,
Z_{JM} = \left( \begin{array}{c}
A_{JM}^z \\ B_{JM}^z \end{array} \right),
\end{equation}
and 
\begin{subequations}\label{Eq:MijJMMulti}
\begin{equation}
M_{JM}^{ij} = \left( \begin{array}{cc}
 0 & \hat{\mathcal{H}}^{(0)}_J-E_i \\
\hat{\mathcal{H}}^{(0)}_J+8\pi G m_0^2\sigma_i^{(0)}\Delta_J^{-1}\left[\sigma_i^{(0)}\right]-E_i & 0  
\end{array} \right)\label{Expresion:Mij}
\end{equation}
for $i=j$, and 
\begin{equation}
M_{JM}^{ij} = \left( \begin{array}{cc}
 0 &0 \\
8\pi G m_0^2\sigma_i^{(0)}\Delta_J^{-1}\left[\sigma_j^{(0)}\right]  &  0
\end{array} \right)\label{Expresion:Mij2}
\end{equation}
\end{subequations}
for $i\neq j$. Notice that we have also introduced the operator:
\begin{subequations}
\begin{align}
\hat{\mathcal{H}}_J^{(0)}&:=-\frac{1}{2m_0}\Delta_J + 4\pi G m_0^2\Delta_s^{-1}\left(\vec{\sigma}^{(0)*}\cdot\vec{\sigma}^{(0)}\right),\label{Eq.OpeH0_j}
\end{align}
with
\begin{eqnarray}\label{Eq:Lap_and_LapJInv}
\Delta_J &:=& \Delta_s - \frac{J(J+1)}{r^2},\\
\Delta_J^{-1}(f)(r) &:=& -\frac{1}{2J+1}\int_0^\infty \frac{r_{<}^{J}}{r_{>}^{J+1}} f(\tilde{r})\tilde{r}^2 d\tilde{r},
\label{Eq:LapJInv}
\end{eqnarray}
\end{subequations}
where $r_< := \min\left\lbrace r,\tilde{r}\right\rbrace$ and $r_> := \max\left\lbrace r,\tilde{r}\right\rbrace$. The action of $\Delta_J^{-1}[\sigma^{(0)}]$ on a function $f$ is defined as $\Delta_J^{-1}[\sigma^{(0)}] f := \Delta_J^{-1}[\sigma^{(0)} f]$. Note that $\Delta_s := \Delta_{J=0} = \tfrac{1}{r}\tfrac{d^2}{dr^2}r$ and $\Delta_s^{-1}:= \Delta_{J=0}^{-1}$ denote the radial Laplacian and its inverse, respectively.

Equations~(\ref{eq.general system}), with suitable boundary conditions (see App.~C in Ref.~\cite{Nambo:2025lnu}), define a linear eigenvalue problem for the spectral parameter $\lambda$. A configuration is said to be mode-stable if all the eigenvalues are purely imaginary. In contrast, the existence of an eigenvalue with a positive real part $\lambda_R>0$ indicates the presence of a linear instability.  Mode-stability does not necessarily imply the stability of the configuration. However, whenever numerical simulations were performed, the evolution was found to be consistent with the stability analysis presented here. The results of these simulations will be presented in a forthcoming paper.
For a complete derivation of the perturbation system and a discussion of its main properties, we refer the reader to Ref.~\cite{Nambo:2025lnu}.

\section{Solution space}\label{sec.solution.space}

In this section, we perform a systematic study of the solution space of spherical multi-frequency Proca stars. Previous work~\cite{Nambo:2024hao, Nambo:2025lnu} has provided a comprehensive analysis of $1$-component multi-frequency states and a preliminary investigation of the $2$-component case. Here, we extend and complement those results. A classification of the solutions based on their stability properties, obtained from a linear perturbation analysis, is presented in Sec.~\ref{sec.stability}.

For the numerical integration of the system~(\ref{eqs.multi.frequency}), we introduce the shifted functions
\begin{align}\label{Eq.shifted_functions}
u_i^{(0)}(r):= E_i - m_0 \mathcal{U}(r),     
\end{align}
where $\mathcal{U}(r):= 4\pi G m_0 \Delta^{-1}_{s} \left(|\vec{\psi}(r)|^2\right)$ denotes the gravitational potential. We also define the following dimensionless quantities:
\begin{subequations}\label{eq.code.numbers1}
\begin{eqnarray}
{\displaystyle t:=4\pi G m_{0}^3\, t^{phys},} &\quad {\displaystyle  \vec{x} := \sqrt{8\pi G} m_{0}^2\, \vec{x}^{phys},} \\
{\displaystyle \mathcal{U} := \frac{1}{4\pi G m_0^2}\,\mathcal{U}^{phys},} &\quad {\displaystyle \vec{\psi}:= \frac{1 }{\sqrt{8\pi G}m_0^{5/2}}\,\vec{\psi}^{phys},} 
\end{eqnarray}
\end{subequations}	
so that the system~(\ref{eqs.multi.frequency}) can be expressed in the more convenient form:
\begin{subequations}\label{eqs.numerical}
    \begin{align}
\Delta_s \sigma_i^{(0)}&= -u_i^{(0)} \sigma_i^{(0)},\label{SEc2.2.2.1}\\
\Delta_s u_i^{(0)}&= - \sum_{j}\sigma_j^{(0)2}.
    \end{align}
\end{subequations}
From this point onward, we will work with dimensionless units unless otherwise stated.

Equations~(\ref{eqs.numerical}) must be complemented with boundary conditions. These are obtained from Eqs.~(\ref{Eq.BounCond.multy.1}), rewritten in terms of the new variables. To ensure regularity at the origin, one must impose
\begin{subequations}\label{Eq.BounCond.multy}
\begin{align}
\sigma_i^{(0)}(r=0)&=\sigma_{i0},\quad \sigma_i^{(0)\prime}(r=0)=0,\label{eq.bc1}\\
u_i^{(0)}(r=0)&=u_{i0},\hspace{0.4cm} u_i^{(0)}{}'(r=0)=0.\label{eq.bc2}
\end{align}
Furthermore, we are interested in solutions that possess finite total energy, which requires
\begin{equation}\label{Eq.boundary_asymp}
\lim\limits_{r\to\infty}\sigma^{(0)}_i(r)=0. 
\end{equation}
\end{subequations}
Here $\sigma_{i0}$ and $u_{i0}$ are constants, and their appropriate values are determined using a procedure similar to that described in Ref.~\cite{Roque:2023sjl} (see also~\cite{Nambo:2024hao, Nambo:2025lnu}). In particular, for a given choice of $\sigma_{i0}$, the admissible values of $u_{i0}$ are fine-tuned via a numerical shooting method, enforcing the asymptotic condition~(\ref{Eq.boundary_asymp}). The asymptotic values of $u_i^{(0)}$ at infinity are left unrestricted and are subsequently used to extract the frequencies $E_i$ via Eq.~(\ref{Eq.shifted_functions}).

Finally, in terms of the dimensionless variables defined in Eq.~(\ref{eq.code.numbers1}), the quantities $E_i$, $N_i$ and $\mathcal{E}$ are computed according to~(see Appendix E of Ref.~\cite{Nambo:2024hao}):
\begin{subequations}\label{Eqs.ENEnergy}
\begin{align}
E_i &= u_{i0}-\sum_{i}\int_{0}^{\infty} \sigma_i^{(0)2}(r) r dr,\label{EqIntEnerg}\\
N_i &= 4\pi \int_{0}^{\infty} \sigma_i^{(0)2}(r) r^{2} dr,\\
\mathcal{E} &=  -4\pi \sum_i \int_0^\infty \frac{1}{2} \sigma_i^{(0)\prime2} r^2 dr.
\end{align}
\end{subequations}
%

\subsection{1-component multi-frequency states}\label{sec.1component}

\begin{figure*}[t]
    \includegraphics[width=\textwidth]{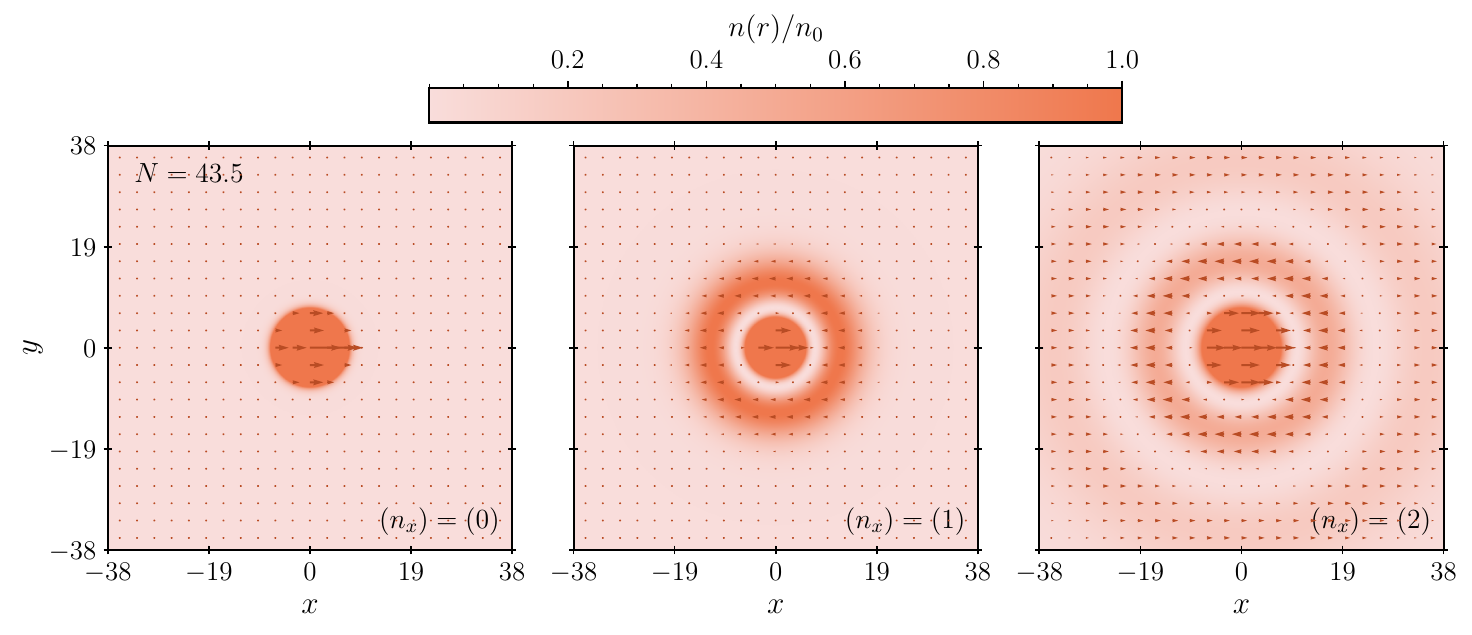}
    \caption{{\bf $1$-component multi-frequency Proca stars (configurations).} Real part of the vector field, $\vec{\psi}_R(t,\vec{x})$ [arrows], and particle number density, $n(t,\vec{x})$ [red shading], at $t=0$ for 1-component multi-frequency states belonging to the families $(n_x)=$ (0), (1) and (2), for $N=43.5$. Each family consists of a single, stationary configuration. The $(n_x)=(0)$ case corresponds to the ground state at fixed $N$. For $(n_x)=$ (1) and (2) the field orientation differs from one shell to another.} \label{Fig.sol_1component}
\end{figure*}

We begin by considering  solutions that oscillate with a single frequency, for which only one component of the vector field $\vec{\psi}(t,\vec{x})$ is nonvanishing.\footnote{If more than one component were nonvanishing, they would necessarily share the same spatial profile (up to a constant proportionality factor). In that case, a global rotation in internal space can always be performed to align the configuration along a single direction, reducing it to a 1-component form~\cite{Nambo:2024hao}.\label{footnote}} Without loss of generality, we choose this to be the $x$-component, so that $\sigma_y^{(0)}(r)=\sigma_z^{(0)}(r)=0$. Under this assumption, the indices $i,j$ in Eqs.~(\ref{eqs.numerical}) are restricted to the single value $x$, and the node number $n_x$ is arbitrary.  
These solutions are equivalent to stationary, linearly polarized Proca stars, which have been extensively studied in the literature; see, e.g., Refs.~\cite{Jain:2021pnk, Zhang:2021xxa, Adshead:2021kvl}.\footnote{Leaving  polarization aside, they are also equivalent to standard $s=0$ boson stars.} Following the notation of this paper, we refer the reader to Refs.~\cite{Nambo:2024hao, Nambo:2025lnu}, where the main properties of these solutions were reported, and where the effects of the particle-particle and spin-spin selfinteractions were also explored.

For a fixed node number, there is only one way to accommodate $N$ particles in a single field component, so the spectrum of 1-component multi-frequency states is discrete, with only one  solution in each $(n_x)$ family. Among them, the case  $(n_x)=0$ is particularly interesting, as it constitutes the ground state of the $s=1$ Schr\"odinger-Poisson system at fixed $N$~\cite{Nambo:2024hao}.  Figure~\ref{Fig.sol_1component} shows 1-component multi-frequency states belonging to the first three families, $(n_x)=$ (0), (1) and (2), for $N=43.5$. This value has no special significance and was chosen simply so that  $(n_x)=(0)$ has unit amplitude, $\sigma_{x0}=1$. Solutions with any other possible value of $N$ can be obtained from these configurations using the scaling~(\ref{eq.scaling.free}). 

\subsection{2-component multi-frequency states}\label{sec.2component}
\begin{figure}[t]
    \includegraphics[width=\columnwidth]{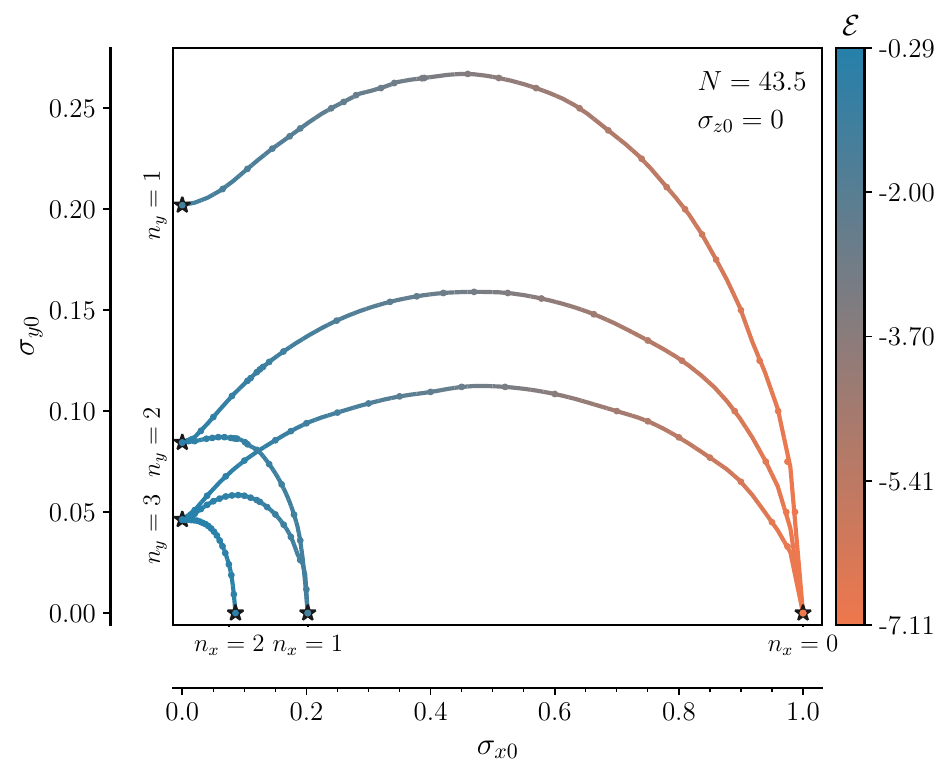}
    \caption{{\bf $2$-component multi-frequency Proca stars.}
    2-component multi-frequency states belonging to the families $(n_x,n_y)=$ (0,1), (0,2), (0,3), (1,2), (1,3), (2,3), for $N=43.5$. Each family continuously connects the corresponding $(n_x)$ and $(n_y)$ 1-component states. Note that the ground state, $(n_x)=(0)$, is the lowest energy configuration.    Figure~\ref{Fig.sol_2component2} shows representative  configurations from these families.}\label{Fig.sol_space2}
\end{figure}

\begin{figure*}[t]
    \includegraphics[width=\textwidth]{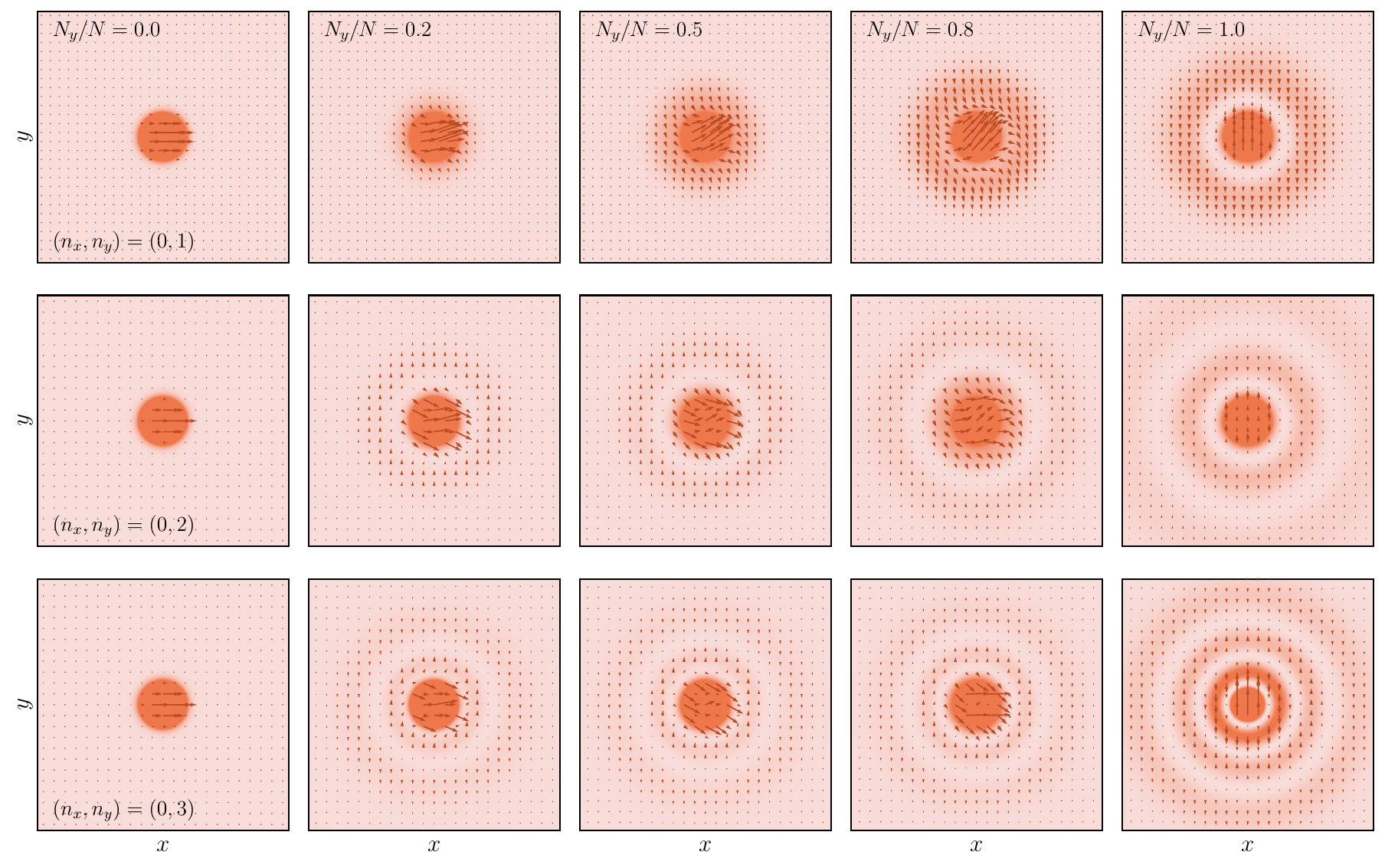}
    \caption{{\bf $2$-component multi-frequency Proca stars (configurations).} Same as Fig.~\ref{Fig.sol_1component}, but for representative configurations from the families $(n_x,n_y)=$ (0,1), (0,2) and (0,3), for $N=43.5$. All these families start from the ground state $(n_x)=0$ and end at  the excited $(n_y=1)$, $(n_y=2)$, and $(n_y=3)$ states, respectively. From left to right, particles are progressively transferred from the $x$ to the $y$ component. A movie illustrating the time evolution of a 2-component multi-frequency Proca star is provided in~\cite{youtube}.}\label{Fig.sol_2component2}
\end{figure*}

We now turn to solutions that oscillate with two distinct frequencies, for which two components of the wave function $\vec{\psi}(t,\vec{x})$ are nonvanishing. Without loss of generality, we take  them to be the $x$- and $y$-components, so that $\sigma_{z}^{(0)}(r)=0$. In this case, the indices $i, j$ in Eqs.~(\ref{eqs.numerical}) run over $x$ and $y$, and the node numbers $n_x$ and $n_y$ are arbitrary, as long as $n_x<n_y$. 

For fixed node numbers, there are multiple ways to distribute $N$ particles between two field components, and therefore the spectrum of 2-component multi-frequency states is continuous, with a one-parameter family of solutions for each $(n_x,n_y)$. Furthermore, each family interpolates between the corresponding 1-component states $(n_x)$ [with $N_x=N$, $N_y=0$] and $(n_y)$ [with $N_x=0$, $N_y=N$] as particles are continuously redistributed between the $\sigma_x^{(0)}(r)$ and $\sigma_y^{(0)}(r)$ components. Figure~\ref{Fig.sol_space2} shows multi-frequency states $(\sigma_{x0},\sigma_{y0})$ belonging to the families $(n_x, n_y)=$ $(0,1)$, $(0,2)$, $(0,3)$, $(1,2)$, $(1,3)$, $(2,3)$, for $N=43.5$, where solutions for any other value of $N$ can be obtained via the scaling~(\ref{eq.scaling.free}). Note that the curve $(0,1)$ was already presented in Fig.~15 of Ref.~\cite{Nambo:2024hao}, and in Fig.~3 of Ref.~\cite{Nambo:2025lnu} for $N=25.4$. For completeness, Fig.~\ref{Fig.sol_2component2} shows some representative  configurations from the first three families as the number of particles is transferred from $N_x$ to $N_y$. As $N_y$ increases, the nodes of the $y$-component become more pronounced, and the arrows gradually change their orientation.\footnote{To properly interpret these figures, it is important to emphasize the following: the particle number density does not exhibit nodes; rather, they display regions of lower energy, which do not exactly coincide with the nodes of the $y$-component. Consequently, in some cases, the arrows point downward before reaching the first low-density region.}

 It is worth noting that multi-frequency states are not critical points of the energy functional at fixed particle number $N$. Instead, they correspond to stationary points under variations that leave the Hermitian second-rank tensor $Q_{ij}=N_i\delta_{ij}$ unchanged~\cite{Nambo:2024hao}. This underlies the existence of continuous families of multi-frequency states, and explains why these solutions can be continuously deformed (by redistributing particles among the different components of the field) while keeping $N$ fixed, thereby increasing or decreasing the energy, as shown in Fig~\ref{Fig.sol_space2}.

\subsection{3-component multi-frequency states}\label{sec.3component}
\begin{figure*}[t]
    \includegraphics[width=\textwidth]{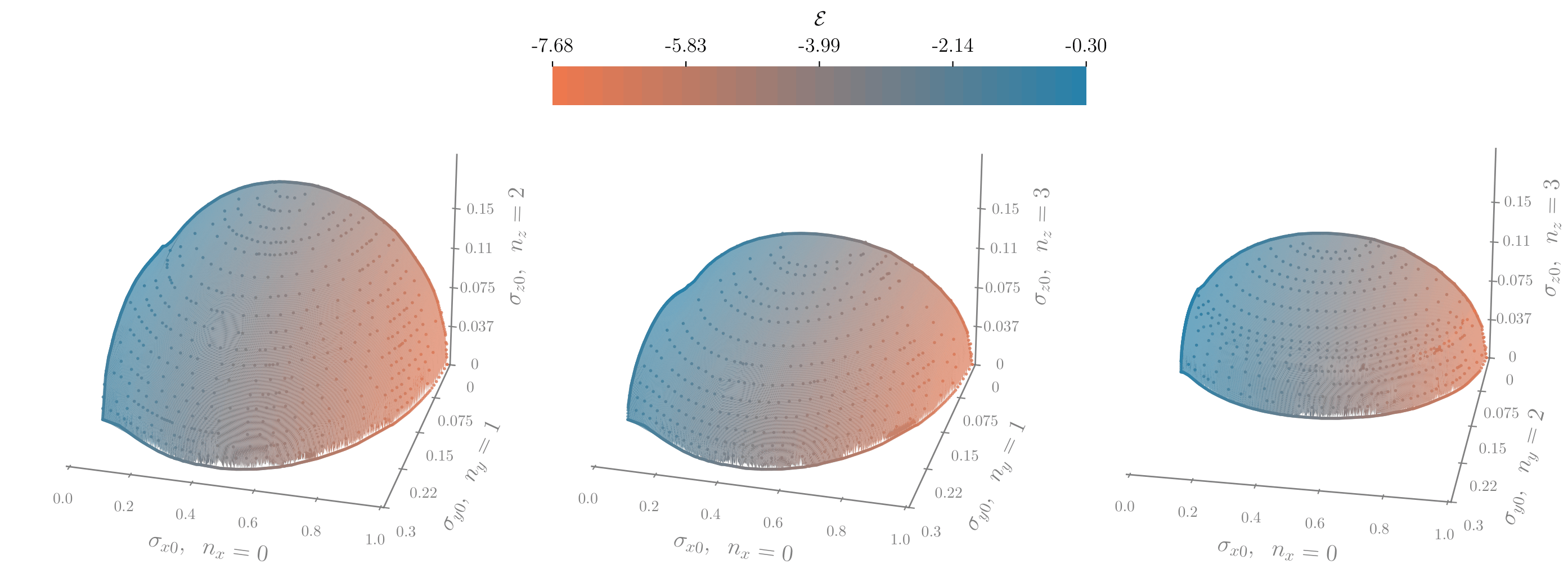}
    \caption{{\bf $3$-component multi-frequency Proca stars.} 3-component multi-frequency states belonging to the families $(n_x,n_y,n_z)=$ (0,1,2), (0,1,3) and (0,2,3), for $N=43.5$. The planes $\sigma_{z0}=0$, $\sigma_{y0}=0$ and $\sigma_{x0}=0$ in these figures correspond to 2-component curves shown in Fig.~\ref{Fig.sol_space2}. Note that the ground state, $(n_x)=(0)$, remains as the lowest energy configuration.}\label{Fig.sol_space3}
\end{figure*}

Finally, we explore the general case of solutions oscillating with three different frequencies, for which all components of the wave function $\vec{\psi}(t,\vec{x})$ are nonvanishing. This constitutes the main focus of the present work. Since we are only interested in stable equilibrium configurations, we will fix $n_x=0$ (see the next section) and consider only configurations with $0<n_y<n_z$.

As in the 2-component case, the spectrum of 3-component multi-frequency states is continuous, forming a two-parameter family of solutions for each $(0,n_y,n_z)$. Each such family defines a two-dimensional surface in the parameter space $(\sigma_{x0},\sigma_{y0},\sigma_{z0})$, with boundary curves defined by the 2-component families $(0,n_y)$, $(0,n_z)$ and $(n_y,n_z)$, which are recovered in the limits where one field component becomes unoccupied as particles are continuously redistributed among the different components. Figure~\ref{Fig.sol_space3} shows multi-frequency states $(\sigma_{x0},\sigma_{y0},\sigma_{z0})$ belonging to the families $(n_x,n_y,n_z)=(0,1,2)$, $(0,1,3)$ and $(0,2,3)$, for $N=43.5$, which can be generalized to arbitrary $N$ using the scaling~(\ref{eq.scaling.free}). In App.~\ref{App:Numerical procedure}, we provide technical details on the numerical procedure used to obtain these solutions. Each dot in the figure represents an individual solution constructed numerically. The color scale, shown in the upper bar and corresponding to the energy, is obtained through interpolation.

Note that, in all cases, the lower energy configurations are located on the right-hand side of the diagrams, where most particles occupy the nodeless component $n_x=0$, with the ground state appearing at the far right. Interestingly, as we show in the next section, stability is not restricted solely to these regions.

\section{Linear stability}\label{sec.stability}

In this section, we study the linear stability of multi-frequency states. To this end, we systematically analyze the eigenvalue spectrum of the linear system~(\ref{eq.general system}) for the configurations reported in the previous section.

In terms of the dimensionless quantities introduced in Eqs.~(\ref{eq.code.numbers1}), the matrices~$M^{ij}_{JM}$ given in~(\ref{Expresion:Mij}) and~(\ref{Expresion:Mij2}) take the form
\begin{subequations}\label{Eq:MijJMMulti_Ad}
\begin{equation}
M_{JM}^{ij} = \left( \begin{array}{cc}
 0 & \hat{\mathcal{H}}^{(0)}_J-E_i \\
\hat{\mathcal{H}}^{(0)}_J+2\sigma_i^{(0)}\Delta_J^{-1}\left[\sigma_i^{(0)}\right]-E_i & 0  
\end{array} \right)
\end{equation}
and
\begin{equation}
M_{JM}^{ij} = \left( \begin{array}{cc}
 0 &0 \\
2\sigma_i^{(0)}\Delta_J^{-1}\left[\sigma_j^{(0)}\right]  &  0
\end{array} \right),
\end{equation}
\end{subequations}
respectively, with 
\begin{equation}
\hat{\mathcal{H}}_J^{(0)}:=-\Delta_J  + \Delta_s^{-1}\left(\vec{\sigma}^{(0)*}\cdot\vec{\sigma}^{(0)}\right)
\end{equation}
and $\Delta_s^{-1}:= \Delta_{J=0}^{-1}$, where $\Delta_J$ and $\Delta_J^{-1}$ take the same form as in Eqs.~(\ref{Eq:Lap_and_LapJInv}) replacing physical by dimensionless quantities. 

For the numerical implementation and discretization of the system~(\ref{eq.general system}), we have followed the procedure described in Appendix~C of Ref.~\cite{Nambo:2025lnu}. In the present work, we focus on the presentation of the results and refer the reader to the cited reference for a comprehensive description of the numerical methodology. A discussion of the strategy used to identify numerical zeros in the 3-component case is presented in App.~\ref{App:Numerical procedure}.

\subsection{Stability of 1-component multi-frequency states}

As pointed out above, 1-component multi-frequency states are equivalent to  stationary, linearly polarized Proca stars, whose linear stability was studied in~Ref.~\cite{Nambo:2025lnu}. In that case, only the ground state $(n_x)=0$ is linearly stable. We therefore do not repeat the analysis here and instead invite the reader to consult the original reference, Sec.~VA$\,$1, for details.

\subsection{Stability of 2-component multi-frequency states}

\begin{figure*}[t]
    \includegraphics[width=\textwidth]{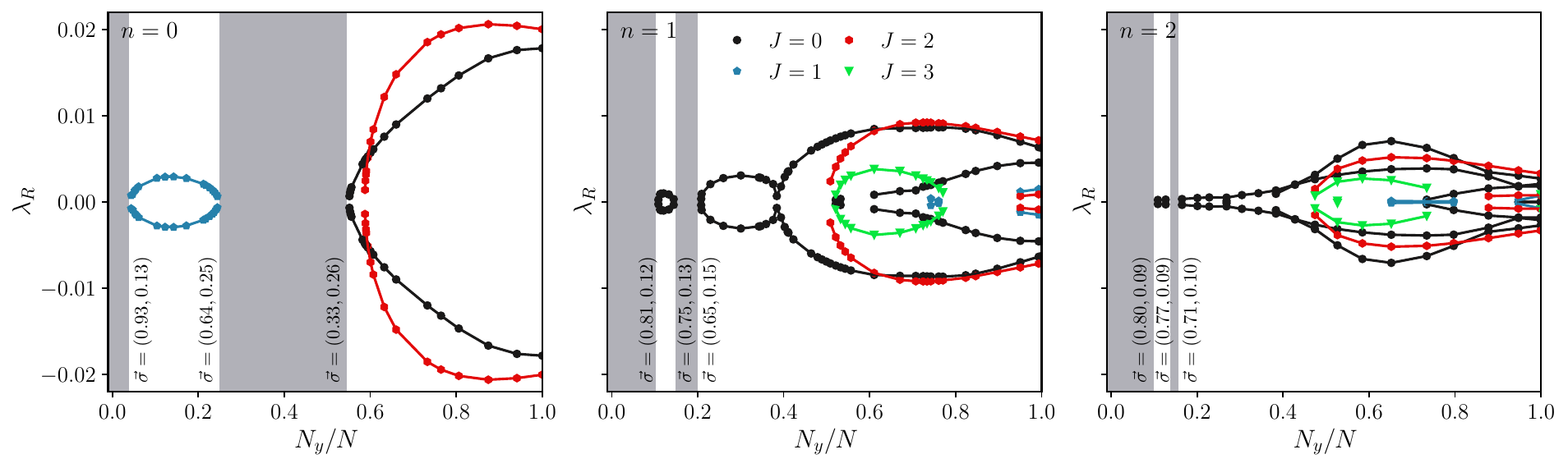}
    \caption{{\bf Real eigenvalue spectra of 2-component multi-frequency Proca stars.} Spectra of the 2-component families $(n_x,n_y)=$ $(0,1)$ [left panel], $(0,2)$ [central panel], and $(0,3)$ [right panel] presented in Sec.~\ref{sec.2component}, as function of the particle number ratio in the $y$-component, for perturbations with angular momentum number $J \le 6$. Modes with a positive real part of the eigenfrequency, $\lambda_R> 0$, signal  the onset of a linear instability. The shaded regions indicate stability bands, where no unstable modes are observed. Families with $n_x\neq 0$ are unstable and are not shown here. }\label{Fig.2.component.stability1}
\end{figure*}

\begin{figure}[t]
    \includegraphics[width=\columnwidth]{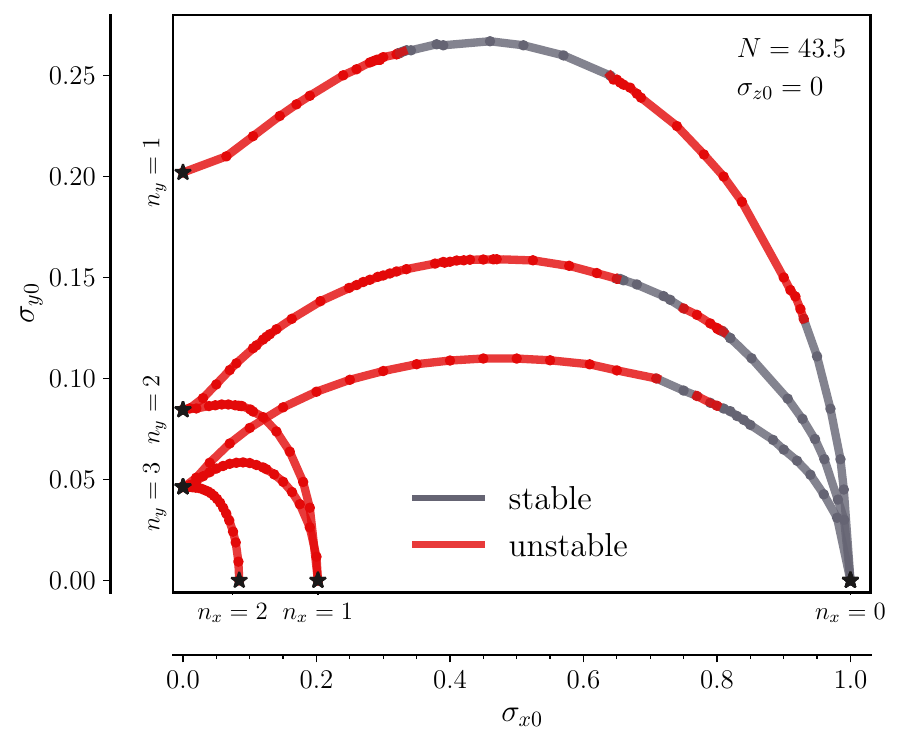}
    \caption{{\bf Linear stability of $2$-component multi-frequency Proca stars.} Same as Fig.~\ref{Fig.sol_space2}, but with stable and unstable states indicated separately. Note that only a subset of states within the $(0,n_y)$ families is linearly stable. All these families start from the ground state.}\label{Fig.2.component.stability2}
\end{figure}

We begin by analyzing the stability of those families that include the ground state $(n_x)=(0)$ as one of their configurations. Figure~\ref{Fig.2.component.stability1} shows the spectrum of the eigenvalue problem~(\ref{eq.general system}) for the 2-component families $(n_x,n_y)=$ $(0,1)$, $(0,2)$ and $(0,3)$. Specifically, we plot the real part of the eigenvalues $\lambda$ as a function of the fraction of particles in the $y$-component, which allows us to identify the onset of unstable modes as the one-parameter family is followed toward the excited $(n_y)$ state. This analysis was performed for values of the angular momentum number up to $J = 6$. Within this range, unstable modes were found only for $J \leq 3$, while no instabilities were observed for higher angular momenta.\footnote{This suggests that potential instabilities are restricted to the low $J$ sector, although the appearance of unstable modes at higher $J$ cannot be ruled out at this stage. Nevertheless, in many selfgravitating systems, unstable modes are typically absent beyond a sufficiently large angular momentum, which in the present case may be as small as $J=3$ (see, for example, Sec.~IV$\,$E of Ref.~\cite{Nambo:2023yut} for a discussion in the context of $\ell$-boson stars). Furthermore, whenever dynamical simulations were carried out for a given configuration, the resulting evolution confirmed the predictions of the linear stability analysis presented here.} These results are consistent with those reported in~Sec.~VA$\,$3 of Ref.~\cite{Nambo:2025lnu}, where the stability of the $(n_x,n_y)=(0,1)$ family was investigated for $N=25.4$. In all cases, the ground state is found to be linearly stable, along with a finite region of nearby configurations, leading to a stability band. However, as one moves towards the opposite end of each family, the solutions eventually become unstable. Moreover, for the three cases we have examined, a second stability band emerges that is not continuously connected to the ground state. Depending on the specific family, the unstable region that separates the two stability bands is associated with different unstable modes.\footnote{It is worth noting that, for the $(n_x,n_y)=$ $(0,2)$ and $(0,3)$ families, the boundaries of the stable regions are determined by unstable $J=0$ modes only. In contrast, for the $(n_x,n_y)=(0,1)$ family, unstable $J=1$ modes also play a role in determining the stability bands, excluding regions of the parameter space that would otherwise be stable against radial perturbations.} As expected, the stable regions decrease their size as the number of nodes $n_y$ increases.
 
For families that do not include the ground state, namely those with $n_x\neq 0$, we repeated the same linear stability analysis for $(n_x,n_y)=$ $(1,2)$, $(1,3)$, $(2,3)$ and found no evidence of linearly stable solutions. In all cases, the configurations at both ends of each family are known to be unstable, and no stable states appear in the intermediate region. This suggests that the presence of the ground state $(n_x)=(0)$ is a necessary condition for stability in 2-component multi-frequency families, as expected.

Figure~\ref{Fig.2.component.stability2} summarizes the results of this section, where the parameter space of the 2-component families presented in Fig.~\ref{Fig.sol_space2} is shown using different colors to distinguish between stable and unstable solutions. From this figure, it is evident that all linearly stable states belong to families that originate from the ground state of the system. Stable multi-frequency states are expected to correspond to local or global minima of the energy functional under variations that maintain $\hat{Q}$ fixed, although a rigorous proof is still lacking.

\begin{figure*}[t]
    \includegraphics[width=\textwidth]{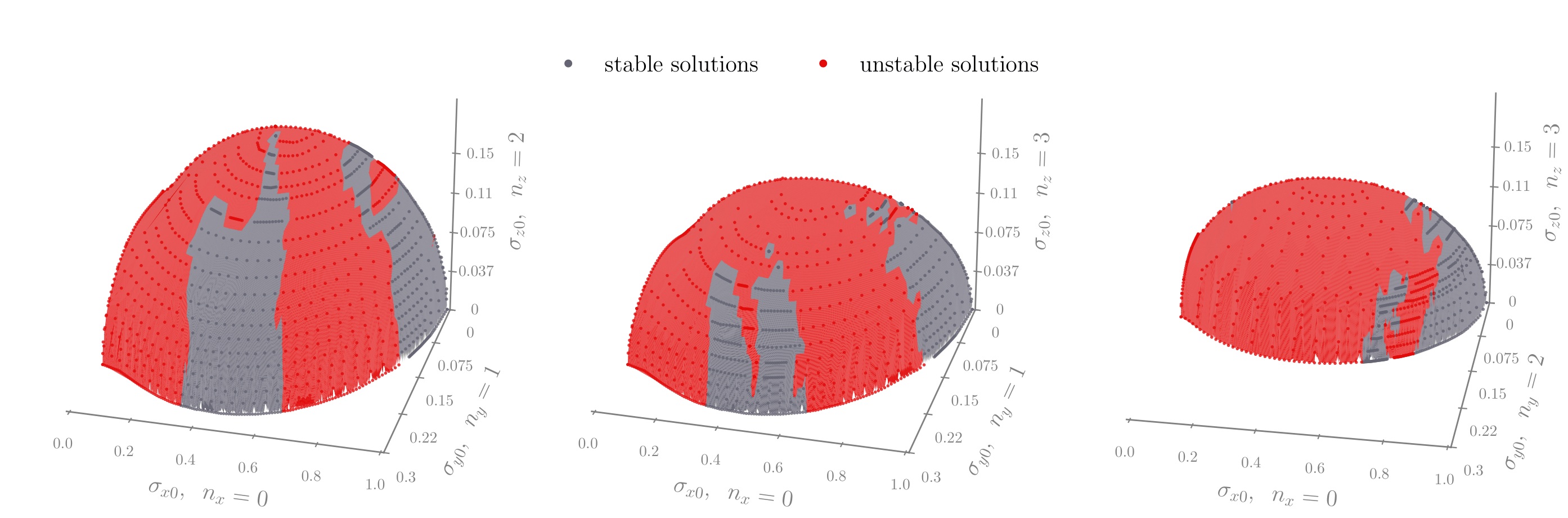}
    \caption{{\bf Linear stability of 3-component multi-frequency Proca stars.} Same as Fig.~\ref{Fig.sol_space3}, but with stable and unstable states indicated separately. Only a subset of states within the $(0,n_x,n_y)$ families  are linearly stable. All these families start from the ground state.}
     \label{Fig.stability_regions3D(0,1,2)}
\end{figure*}

\subsection{Stability of 3-component multi-frequency states}

We extend the analysis of the previous subsection to the 3-component case. Since no linearly stable configurations were found in 2-component families that do not include the ground state, we restrict our study to 3-component families of the form $(0,n_y,n_z)$.\footnote{All tests performed for families $(n_x,n_y,n_z)$ with $n_x\neq 0$ have revealed no evidence of stable states.}

Figure~\ref{Fig.stability_regions3D(0,1,2)} shows the parameter space of the $3$-component families introduced in Fig.~\ref{Fig.sol_space3}. As in Fig.~\ref{Fig.2.component.stability2}, stable states are shown in gray, while unstable states in red. In all cases, the two-dimensional solution surface is partitioned into distinct stability regions, and its intersections with the $\sigma_{x0}$, $\sigma_{y0}$  and $\sigma_{z0}$ coordinate planes reproduce the curves reported in Fig.~\ref{Fig.2.component.stability2}, reflecting the continuous reduction of the $3$-component families to their corresponding $2$-component limits as one component of the wave function is depopulated.   

For the $(n_x,n_y,n_z) = (0,1,2)$ family, we identify two different stable regions. The rightmost region (connected to the ground state), gradually decreases in size as the amplitude of the $z$-component is increased, until $\sigma_{z0} \approx 0.10$, where it grows again for a short interval. As the family reaches the first unstable region along the $(n_x,n_y) = (0,2)$ curve, at $\sigma_{z0}\approx 0.12 $, this stable region rapidly shrinks and becomes disconnected from that curve. It reconnects only upon reaching the second stability band, at $\sigma_{z0}\approx 0.14$, and ultimately disappears entirely as this second band terminates at $\sigma_{z0}\approx 0.15$.

For the leftmost stable region, its size remains relatively constant until an instability associated with $J = 0$ modes appears, dividing the region into two around $\sigma_{z0} \approx 0.13$. The rightmost of these subregions eventually disappears as this new unstable band grows in size with increasing $\sigma^{(0)}_{z0}$, approximately when $\sigma_{z0}\approx0.14$, while the left subregion continues only over a very short interval.

A similar behavior is observed for the other families explored. However, as expected, the extent of the stable regions decreases as the number of nodes $n_y$ and $n_z$ increases. At the same time, these configurations become more difficult to construct due to their growing number of nodes, making higher-node families increasingly challenging to analyze in detail. For this reason, and because these families exhibit fewer stable configurations, we have restricted our study to the first three families only.

As we anticipated, stable configurations are not restricted only to the low-energy region, located on the right-hand side of each panel, close to the ground state. Our results indicate that, although stability is generally favored when the majority of the particles occupy the nodeless component rather than the components with one or more nodes, this condition is not strictly necessary. Indeed, stable configurations exist even when this hierarchy is not satisfied. For instance, in the family $(n_x,n_y)=(0,1)$, the second stability band extends up to $N_y/N\approx 0.55$. Furthermore, as illustrated in Fig.~\ref{Fig.stability_regions3D(0,1,2)} (see also Fig.~\ref{Fig.2.component.stability1}), this criterion alone does not ensure the stability of the configuration, and there exist solutions for which the particles are accommodated in their majority in the nodeless component and are nevertheless unstable.

\section{Conclusions}

In this paper, we have performed a systematic study of the solution space of nonrelativistic multi-frequency Proca stars, including a detailed analysis of their linear stability. Multi-frequency states were first constructed in Ref.~\cite{Nambo:2024hao} (see also~\cite{Nambo:2025lnu}) as classical solutions of the $s=1$ Schr\"odinger-Poisson system, and, given their recent introduction, remain relatively unexplored in the literature. However, the existence of linearly stable multi-frequency states indicates that, in principle, they could form dynamically and persist in virialized systems of spin-1 particles. This possibility motivates a more careful and comprehensive exploration of these solutions, which has been the main purpose of this work.

Multi-frequency states are organized into classes, defined by the number of independent frequencies (or, equivalently, independent field components) simultaneously present in a given configuration. Each class is further subdivided into families, characterized by the node numbers $n_i$ of the individual components of the vector wave function $\vec{\psi}(t,\vec{x})$. These families are labeled by the ordered triple $(n_x,n_y,n_z)$, where the entries can be chosen, without loss of generality, such that $n_x<n_y<n_z$. A summary of this classification for spherically symmetric multi-frequency states, along with their main properties, is provided in Table~\ref{table}.

1-component multi-frequency states $(n_x)$ oscillate with one frequency and involve only one component of the wave function $\vec{\psi}(t,\vec{x})$. They are therefore equivalent to stationary (i.e. single-frequency) states of constant linear polarization. These states have been exhaustively explored in~\cite{Nambo:2024hao,Nambo:2025lnu}, where it was shown that only the ground state $(n_x)=(0)$ is linearly stable. 

2-component multi-frequency states $(n_x,n_y)$ oscillate with two distinct frequencies and involve two components of the wave function $\vec{\psi}(t,\vec{x})$. They were first introduced in Ref.~\cite{Nambo:2024hao}, and their stability was explored preliminarily in Ref.~\cite{Nambo:2025lnu}, where it was shown that some of these configurations are linearly stable. In the present work, we have extended this analysis and found that linear stability is restricted to families of the form $(n_x,n_y)=(0,n_y)$. Moreover, we have identified the stable states within the first three such families, which are distributed along two distinct stability bands in each case. The first band is continuously connected to the ground state, whereas the second is not. For the $(n_x,n_y)=(0,1)$ family, the unstable band separating the two stable regions is triggered by $J=1$ modes, demonstrating the existence of configurations that are stable under radial perturbations but unstable against more general non-radial perturbations.

3-component multi-frequency states $(n_x,n_y,n_z)$ oscillate with three different frequencies and involve all components of the wave function $\vec{\psi}(t,\vec{x})$. This represents the most general realization of multi-frequency configurations and, to our knowledge, the present paper provides the first systematic exploration of such solutions. We have found that linear stability is restricted to families of the form $(n_x,n_y,n_z)=(0,n_y,n_z)$, and have identified the stable states within the lowest such families. In general, stable states lie close to the ground state, although this is not always the case. There exist stable configurations in which a substantial fraction of particles occupy a component with nodes, as well as unstable configurations in which the majority of particles reside in the nodeless component.

The main contribution of the present work is to add further evidence that the phenomenology of selfgravitating spin-1 solitons is significantly richer than in the scalar (spin-0) case, for which stability is restricted to the ground state. Previous studies~\cite{Nambo:2024hao, Nambo:2025lnu} demonstrated that, at fixed particle number $N$, a radially polarized configuration exists in addition to the ground state that is stable against arbitrary linear perturbations. Here, we have extended this picture to multi-frequency configurations, identifying new classes of linearly stable states that complement the previously known stable solutions. Further, our results reveal a qualitatively new structure of the solution space, characterized by continuous families of stable equilibrium configurations.

 For ultralight dark matter candidates~\cite{Marsh:2015xka,Hui:2021tkt,Ferreira:2020fam}, the existence of multi-frequency states may provide new observational signatures of the underlying particle spin in astrophysical data. In pulsar timing, for instance, the coherently oscillating field induces periodic variations in the gravitational potential of the halo, leading to observable timing residuals in scalar~\cite{Khmelnitsky:2013lxt,Porayko:2018sfa}, vector~\cite{Nomura:2019cvc}, and higher-rank tensor field~\cite{Wu:2023dnp} models. In this context, multi-frequency states introduce an additional layer of complexity: as different components of the field oscillate with distinct frequencies, the gravitational potential contains interference terms that give rise to beat phenomena, resulting in a richer and potentially distinguishable structure in the timing signal, including amplitude modulations and multiple spectral features.

Complementary to these time-dependent signatures, the spatial structure of ultralight dark matter halos can also leave observable imprints on galactic dynamics, which is sensitive to the underlying mass density profile of the halo~\cite{Gonzalez-Morales:2016yaf,DeMartino:2018zkx,Zimmermann:2024xvd}. For a fixed boson mass, multi-frequency states are typically more extended than the ground state and exhibit nontrivial internal structure, potentially leaving additional imprints on observables such as rotation curves or stellar kinematics, warranting further investigation.

Another possible source of observable signals, particularly for higher-mass dark matter candidates, arises from direct couplings between the light field and Standard Model particles. Such couplings can induce an effective charge on macroscopic objects under the new interaction, giving rise to dipole radiation in astrophysical systems which, in the presence of mixing with electromagnetism, can be emitted as observable photons~\cite{Amin:2021tnq,Schiappacasse:2026ems}. While existing analyses focus on single-frequency configurations, multi-frequency states are expected to yield a richer phenomenology, including multiple emission channels and interference-induced modulations. Additional observational signatures of ultralight vector dark matter have been investigated in~\cite{LopezNacir:2018epg,Amin:2022pzv, PPTA:2022eul, KAGRA:2024ipf, Dror:2025nvg,Chase:2025wwj,Amaral:2025fcd}, and may be further enriched by the existence of multi-frequency solutions.

In conclusion, the existence and properties of multi-frequency states may provide a novel avenue to probe the particle spin through astrophysical observations, complementing more direct detection strategies based on polarization-dependent effects. These results motivate future dynamical simulations to investigate the formation and survival of multi-frequency states in realistic cosmological and galactic environments, which will be the subject of future work.

\begin{acknowledgments}

We thank Olivier Sarbach for helpful discussions and valuable comments on a previous version of the manuscript. G.D.A. and J.L.M.G. were supported by SECIHTI predoctoral scholarships, and A.A.R. by the postdoctoral fellowship ``Estancias Posdoctorales por México para la Formación y Consolidación de las y los Investigadores por México''. A.D.T. and A.A.R. acknowledge support from SECIHTI-SNII. Additional support was provided by the DAIP project CIIC 301/2026. We also acknowledge the use of the computing server COUGHS from the UGDataLab at the Physics Department of Guanajuato University.

\end{acknowledgments}

\appendix

\section{Numerical procedure for computing and identifying unstable modes}\label{App:Numerical procedure} 

In this Appendix, we describe the numerical procedure used to construct the $3$-component solutions and identify unstable modes. The corresponding analysis for the remaining cases can be found in Ref.~\cite{Nambo:2025lnu}.

\textbf{Background configurations.}
To construct a three-component family of configurations with total particle number $N = 43.5$, satisfying the boundary condition~(\ref{eq.bc1},~\ref{eq.bc2}) and the asymptotic behavior~(\ref{Eq.boundary_asymp}), we extend the methodology introduced in Ref.~\cite{Moroz:1998dh} to the case of two and three components. We first fix the central amplitudes of two components and perform a bijective shooting on the third one. The resulting values of the central amplitudes $u_{i0}$ are then used as initial seeds for a second step, where the nonlinear shooting method described in Ref.~\cite{Singh_2024} is applied.

Due to the numerical precision limitations (approximately $16$ decimal digits in our implementation), the shooting method described above allows us to integrate the system only up to a finite radius. Beyond this point, we employ the asymptotic solutions
\begin{align}\label{Eq.AsymForm}
    \sigma_{i}^{(0)}(r) \approx\frac{C_i}{r} e^{-\sqrt{|E_i|} r}, \quad u_i^{(0)}(r) \approx E_i+\frac{N}{4\pi r}
\end{align}
of Eqs.~(\ref{eqs.numerical}), where $C_i, E_i$, and $N$ denote the amplitude scales, the dimensionless frequencies, and the total particle number, respectively. The value of $E_i$ and $N$ are computed using Eqs.~(\ref{Eqs.ENEnergy}), while the coefficients $C_i$ are determined through a linear fitting procedure.\footnote{Alternatively, the asymptotic form in Eq.~(\ref{Eq.AsymForm}) can be used to extract the eigenvalues $E_i$. This method was employed to validate the results obtained from Eq.~(\ref{EqIntEnerg}).} Further details can be found in Appendix~C of Ref.~\cite{Roque:2023sjl}.

\textbf{Stability.}
The system~\eqref{eq.general system} is solved using the methodology described in Appendix C of Ref.~\cite{Nambo:2025lnu}. Eigenvalues $\lambda$ with nonzero real parts, together with their corresponding eigenvectors, are extracted to identify unstable modes.

During the eigenvalue filtering procedure, spurious solutions associated with the zero mode $\lambda = 0$ may appear due to numerical errors. These modes correspond to perturbations of the form: $(\vec{\mathcal{A}}, \vec{\mathcal{B}}) = \frac{\beta}{2}(- \text{Im}\vec{\sigma}^{(0)},i \text{Re}\, \vec{\sigma}^{(0)})$, where $\beta$ is an arbitrary real constant (see Sec.~IIIA of Ref.~\cite{Nambo:2025lnu} for details on this type of solution). Such spurious modes can be readily identified by inspecting the perturbation $\vec{\sigma}$ [Eq.~\eqref{Eq:PertAnsatz}] and verifying that it vanishes identically for all radii, $\vec{\sigma} = \vec{0}$. When this occurs, the corresponding modes are discarded.

\bibliography{ref.bib} 

\end{document}